\documentclass[oldversion]{aa}     % use v6.0 with [oldversion] or aa_urb.cls without
\def \th {\thinspace}
\def \arcmin {\hbox{$^\prime$}}

\def\approxgt{\mathrel{\hbox{\rlap{\lower.55ex \hbox {$\sim$}} \kern-.3em \raise.4ex \hbox{$>$}}}}
\def\lesssim{\mathrel{\hbox{\rlap{\lower.55ex \hbox {$\sim$}} \kern-.3em \raise.4ex \hbox{$<$}}}}
\def\approxlt{\mathrel{\hbox{\rlap{\lower.55ex \hbox {$\sim$}} \kern-.3em \raise.4ex \hbox{$<$}}}}
\def \degmark {^\circ}

\usepackage{epsf}
\usepackage{graphicx}
\begin{document}

\title{{\bf Neutral absorber dips in the periodic burster LMXB XB\th 1323-619 from Suzaku}}

\author{M. Ba\l uci\'nska-Church\inst{1,2}
\and T. Dotani\inst{3,4}
\and T. Hirotsu\inst{3,4}
\and M. J. Church\inst{1,2}}
\institute{
           School of Physics and Astronomy, University of Birmingham,
           Birmingham, B15 2TT, UK\\
\and
           Astronomical Observatory, Jagiellonian University,
           ul. Orla 171, 30-244 Cracow, Poland.\\
\and     
           Institute of Space \& Astronautical Science,
           3-1-1 Yoshinodai, Sagamihara, Kanagawa 229-8510, Japan.\\
\and       
           Department of Physics, Tokyo Institute of Technology,
           Ohokayama, Meguro, Tokyo, 152-8551, Japan.\\}

\offprints{mbc@star.sr.bham.ac.uk}
%\thanks{}

\date{Received 24 October 2008; Accepted 27 March 2009}
\titlerunning{Suzaku observations of XB\th 1323-619}
\authorrunning{Ba\l uci\'nska-Church et al.}

\abstract{
We present results of an observation with {\it Suzaku} of the dipping, periodic bursting low mass X-ray binary
XB\th 1323-619. Using the energy band 0.8 - 70 keV, we show that the source spectrum is well-described as
the emission of an extended accretion disk corona, plus a small contribution of blackbody emission from the neutron star.
The dip spectrum is well-fitted by the progressive covering model in which the extended ADC is progressively 
overlapped by the absorbing
bulge of low ionization state in the outer accretion disk and that dipping is basically due to photoelectric 
absorption in the bulge. An energy-independent decrease of flux at high energies (20 - 70 keV) is shown to be 
consistent with the level of Thomson scattering expected in the bulge. 
An absorption feature at 6.67 keV (Fe XXV) is detected in the non-dip spectrum and other possible weak features.
In dipping, absorption lines of medium and highly ionized states are seen suggestive of absorption in the ADC
but there is no evidence that the lines are stronger than in non-dip.
We show that the luminosity of the source has changed substantially since the {\it Exosat} observation
of 1985, increasing in luminosity between 1985 and 2003, then in 2003 - 2007 falling to the initial low value.
X-ray bursting has again become periodic,
which it ceased to do in its highest luminosity state, and we find that the X-ray bursts exhibit both the fast
decay and later slow decay characteristic of the rp burning process.
We present arguments against 
the recent proposal that the decrease of continuum flux in the dipping LMXB in general can be explained as 
absorption in an ionized absorber rather than in the bulge in the outer disk generally accepted to be the site 
of absorption.
}
% keywords must be inside abstract
\maketitle

\section{Introduction}

XB\th 1323-619 is a member of the dipping class of Low Mass X-ray Binaries (LMXB), exhibiting reductions 
in X-ray intensity at the orbital period due to absorption in the bulge of the outer accretion disk. 
(White \& Swank 1982; Walter et al. 1982). It is remarkable as one of the two ``clocked'' X-ray bursts 
sources in which bursting remains close to periodic over extended periods of time, the other source 
being 4U\th 1826-24 (Galloway et al. 2004). Other sources such as 4U\th 1705-44 (Langmeier et al. 1987) have 
shown bursting close to periodic, but this is not sustained over long periods of time (Cornelisse et al. 2003).
The dipping sources display relatively complex spectral evolution in dipping, which provides
strong constraints on emission models for LMXB in general since the model must be able to fit the non-dip state
and a sequence of spectra formed by selection in intensity bands at different depths of dipping.
Work over a period of 15 years provided substantial evidence for a model in which the X-ray emission
consists primarily of Comptonized emission from the accretion disk corona (ADC) plus simple
blackbody emission from the neutron star (Church et al. 1997, 1998a,b, 2005; Ba\l uci\'nska-Church et al.
1999, 2000a; Smale et al. 2001; Barnard et al. 2001). This work establishes that in the dip sources in general,
the dominant Comptonized emission is removed slowly showing that the emission is extended,
while the point-like blackbody emission which we associate with the neutron star is rapidly removed, 
and causes rapid variability in dipping. Measurement of dip ingress times allows the size of the 
extended emitter to be measured , and this has shown
the ADC to reach over a large part of the inner accretion disk, have a radial extent $R$ 
between 20\th 000 and 700\th 000 km,
and is thin, with height $H$ having $H/R$ $<<$ 1 (Church \& Ba\l uci\'nska-Church 2004).
Strong independent support for the extended ADC comes from the {\it Chandra} grating results of Schulz et al. (2009)
in which emission lines of highly ionizied species were found to originate at radial distances between
20\th 000 and 110\th 000 km in an extended ADC. Various authors have addressed theoretically the question of the formation,
size and properties of an ADC, such as Jimenez-Garate et al. (2002, and references therein) by considering
the effects of illumination by the central source.

In the present work we investigate spectral evolution in dipping utilising the broadband capability
of {\it Suzaku}, the long-term luminosity evolution of the source and the associated
strong changes in burst properties. In general, the X-ray burst sources
do not have regular bursting. If it is assumed that the mass accretion rate does not vary greatly
on timescales of a few hours, regular bursting would be expected, since mass accumulates at a steady
rate on the surface of the neutron star and conditions for unstable nuclear burning are achieved.
However, the majority of sources have quite irregular bursting and the study of the two ``clocked
bursters'' is clearly of interest in explaining this. The source was observed by us using {\it XMM-Newton}
in 2003, January, and analysis of this observation and comparison with observations previously made with
{\it Exosat}, {\it ASCA}, and three observations with {\it RXTE} revealed relatively large systematic 
changes in burst properties and in the luminosity of XB\th 1323-619 over the period 1985 - 2003 
(Church et al. 2005). The most striking feature was the changing burst rate,
the time between bursts  $\Delta t$ decreasing linearly with time from 322 minutes in 1985 
to 58 min in 2003. During this period the X-ray luminosity in the band 1 - 10 keV increased substantially
from $1.3\times 10^{36}$ to $5.4\times 10^{36}$ erg s$^{-1}$. The observed trend if continued would mean that
the burst rate would continue to increase, implying that $\Delta t$ would become zero on Jan 11, 2008.
However, analysis of a further observation with {\it RXTE} in 2003, September showed 
a sudden change in the source, the luminosity decreasing and the time between bursts increasing 
to a mean of 162 min (Ba\l uci\'nska-Church et al. 2008). 

XB\th 1323-619 was discovered using {\it Uhuru} and {\it Ariel~V} (Forman et al. 1978; Warwick et al. 1981)
and X-ray bursting and dipping found using {\it Exosat} by van der Klis et al. (1985). 
An orbital period of 2.938$\pm$ 0.020 hr was determined by Ba\l uci\'nska-Church et al. (1999) in good agreement
with the value of 2.932$\pm$ 0.005 hr from {\it Exosat} (Parmar et al. 1989).
The source was observed with {\it BeppoSAX} in 1997 from which a detailed study of the broadband 1 - 200 keV 
spectrum was made (Ba\l uci\'nska-Church et al. 1999) showing the Comptonized emission to have a relatively high 
cut-off energy of 44 keV. The {\it XMM} observation (Church et al. 2005) revealed various line and possible edge 
features, including Fe absorption lines at 6.70 and 6.98 keV equivalent to
Fe XXV K$_{\alpha}$ and Fe XXVI K$_{\alpha}$ and an emission line at 6.82 keV. In addition absorption features 
were seen at 1.46 keV and 4.05 keV, possibly Mg XII and Ca XX. The origin of lines was discussed and
it was concluded that the evidence favoured production in the ADC.

In analysis of the {\it XMM} data, Boirin et al. (2005) and D\' iaz-Trigo (2006) suggested that the dipping 
in all LMXB dipping sources could be explained in terms of an ionized absorber rather than in the 
low ionization state bulge in the outer disk generally accepted as the site of absorption. This model
would require dipping to be unconnected with the bulge in the outer disk.
In the present observation we show that dipping is very well explained as absorption in the bulge
plus a small degree of energy-independent electron scattering consistent with the measured column
density of the cool absorber. In Sect. 4 we discuss general objections to the ionized absorber model.

\section{Observations and analysis}

We observed XB\th 1323-619 with {\it Suzaku} in 2007, Jan 9 - 10 for 34.1 
hours, this covering 11.7 orbital cycles. Data from both the XIS and HXD 
instruments were used. Analysis was carried out using the {\it Suzaku}-specific
{\it Ftools}, part of {\sc ftools 6.3.1}. At ISAS, the XIS raw telemetry files were 
converted to Fits format using the facility {\sc mk1stfits}, and then the raw data files were 
converted to filtered and calibrated events files using {\sc mk2ndfits}. This second stage 
was carried using the improvements in calibration data of August, 2007 (version 2 processing). 
The XIS was operated in the normal mode, using the one-quarter window option viewing 1/4 of the CCD
and with the 3x3 and 5x5 pixel options for pulse height analysis and with 2 s 
time resolution. Spaced-row charge injection (SCI) was used to reduce the effects of
radiation damage.
Data for each of the three detectors XIS0, XIS1 and XIS3 detectors were available, 
XIS2 having failed prior to this observation. The filtering consisted of the removal of hot and 
flickering pixels and of selection on event ``grade'' from the pulse height distribution (keeping
grades 0, 2, 3, 4 and 6) to discriminate between X-ray and charged particle events, and also standard
screening based on housekeeping data as follows. 

This standard screening excluded periods of South Atlantic Anomaly passage plus an 
elapsed time of 436 seconds after the time of leaving SAA (T\_SAA $>$ 436), to select data for 
elevation above the Earth's limb more than 5 degrees (ELV $>$ 5), for elevation above the sunlit Earth 
more than 20 degrees (DYE\_ELV $>$ 20), and for geomagnetic rigidity (COR $>$ 6). Lightcurves were
extracted from each XIS detector from the filtered events file using {\it Xselect}. The lightcurve obtained after screening 
revealed only a single burst and 9 dips, whereas a lightcurve made with relaxed screening criteria (below)
showed 5 bursts, the last of which was double, and eleven dips, so clearly standard screening removed 
substantial amounts of data of interest and we investigated the effects of relaxing the selection criteria.

To do this, it was necessary to use the unfiltered events file. Firstly, hot and flickering pixels
were removed using the {\sc ftool} {\sc cleansis} and grade selection was made within {\it Xselect}.
Inspection of the housekeeping data showed that during the observation, elevation  of the source above 
the Earth's limb was always above 7.5$\degmark$, i.e. no Earth occultation took place and so there was no 
need to screen for this. However, we found that when we reduced the daytime elevation selection criterion
to DYE\_ELV $>$ \hbox{10 $\degmark$,} more bursts were revealed and we eventually made no selection
on this parameter. We then checked the effect of this on spectral data, since X-rays from the bright Earth 
have a very soft spectrum dominated by O and N lines below 1 keV. Non-burst spectra with and without screening 
for DYE\_ELV were extracted, but no differences between them could be seen, particularly below 1 keV. As the
spectrum of XB\th 1323-619 in the XIS does not extend below $\sim$0.8 keV because of its low Galactic 
latitude source and associated large column density, little effect would be expected.

The geomagnetic rigidity (COR) was also examined. With the standard selection: COR $>$ 6, a small fraction of 
data was removed including the first burst at time $\sim $6,000 s (see Fig.1) when the COR value was $\sim$5. 
Therefore to include all the bursts we relaxed the selection criterion 
\begin{figure*}[!th]                                                          % Fig. 1
\begin{center}                                                                % uses 3x3, 5x5 obc modes added
\includegraphics[width=64mm,height=160mm,angle=270]{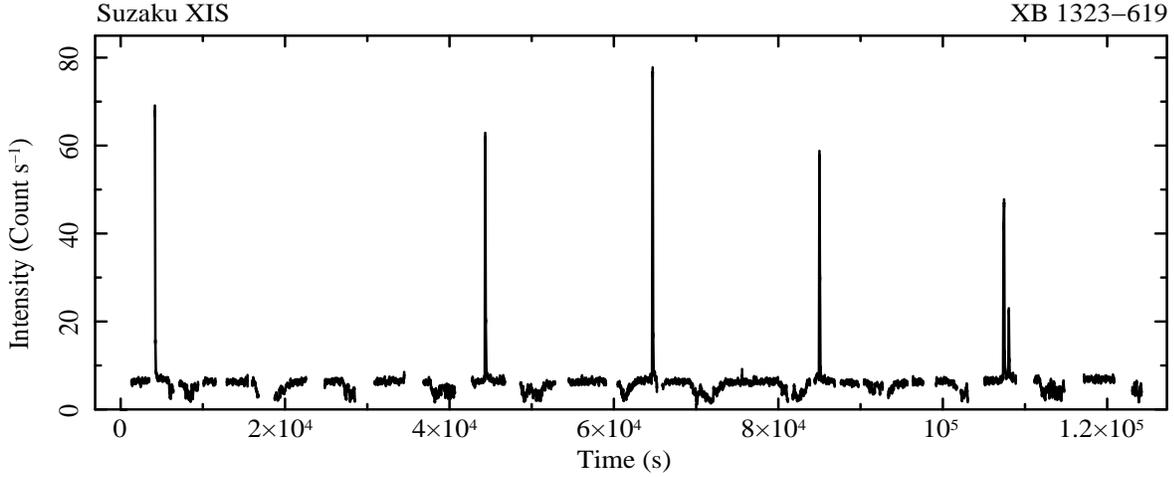}  % was lc_xis_35mm_64s_ff_paper - no no added
\caption{Background-subtracted lightcurve of the January 2007 observation of XB\th 1323-619in the energy band
0.2 - 12 keV with 64 s binning from the XIS detectors XIS0, XIS1 and XIS3}
\end{center}
\end{figure*}
to COR $>$ 5. This may slightly increase 
the background, however as XB\th 1323-619 is a bright source in XIS, the overall effect will be negligible.
It is clear that our approach of relaxing the selection criteria will not affect the determination from the 
lightcurve of the burst repetition rate and decay times, 
%and $\alpha$ parameter 
and our testing showed that the effect on spectra was negligible. 

With the final selections above, the lightcurves from each XIS detector were examined for traces of the nearby 
pulsar 1SAX\th J1324.4-6200  (Angelini et al. 1998; Ba\l uci\'nska-Church et al. 1999) located 17\arcmin 
from XB\th 1323-619 which would have contaminate lightcurves and spectra. The observation had been designed
with a pointing direction to minimize possible effects. No trace of pulsations 
could be seen. A lightcurve of XB\th 1323-619 was then extracted using all 3 XIS detectors which were each
operated in the mode in which one quarter of the full detector was used consisting of a strip of 1024 x 256 pixels.
The lightcurve was extracted in the full energy band 0.2 - 12 keV by specifying an approximately rectangular 
source region with the source at the centre of size 4.4\arcmin x 6.7\arcmin; background was extracted using two square 
regions in the image, each side of the source at a distance of 7$\arcmin$ and 4.4\arcmin square, and adding these. 
Background subtraction was carried out using {\sc lcmath} allowing for the different areas of the source and background 
regions. Photon pileup is negligible. The background-subtracted lightcurve of the 3 XIS detectors is shown in Fig. 1.

\begin{figure*}[!ht]                                                                         % Fig. 2
\begin{center}                            
\includegraphics[width=78mm,height=89mm,angle=270]{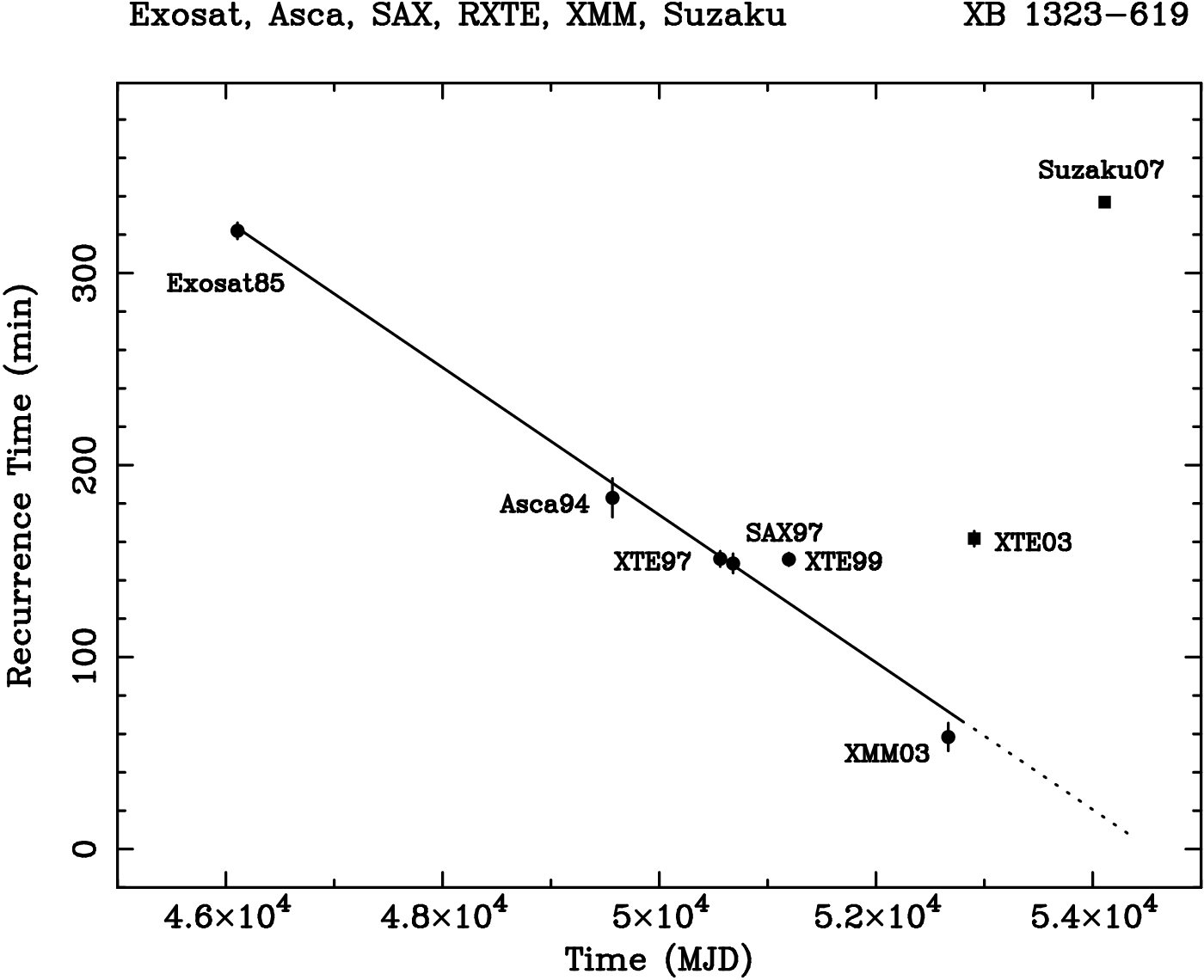}                                % was deltata_suzaku
\includegraphics[width=78mm,height=89mm,angle=270]{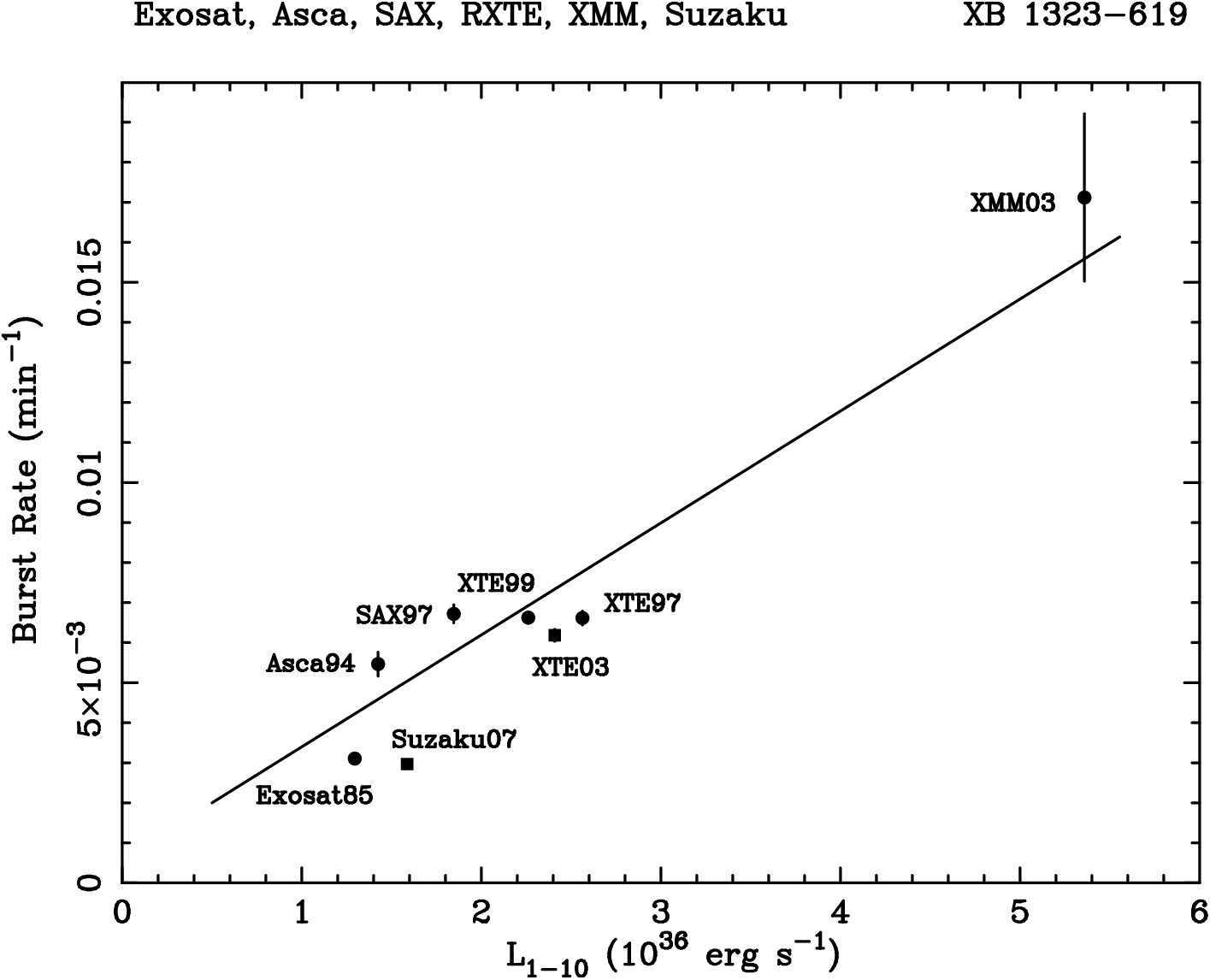}                                % was rate_suzaku
\caption{Left panel: recurrence time $\Delta$t of X-ray bursts showing the 
systematic decrease of $\Delta$t from the {\it Exosat} observation to the 
{\it XMM} observation followed by the departures from that pattern of the 2003 
{\it RXTE} and the present observations; right panel: the dependence of burst rate 1/$\Delta$t as a function of luminosity. 
There is an approximately linear relation from the {\it Exosat} to the {\it XMM} observation, although the 2003 
{\it RXTE} and {\it Suzaku} points have reverted to the positions consistent 
with the much decreased luminosities} 
\end{center}
\end{figure*}

HXD data were obtained in the form of a cleaned events file in which standard screening was carried out for
SAA passage, elevation etc. We analysed the PIN data and found that the source is detected up to $\sim$50 keV. 
Because of this and the weakness of the source, we did not use the high energy GSO data in the band 30 - 600 keV.
Non X-ray background in HXD/PIN is evaluated using a background model provided for observers for each observation.
The cosmic X-ray background which is about 5\% of the total is included as a model term in spectral fitting 
using a power law form corresponding to previous determinations of the spectrum. Data were selected via a good time
interval file (GTI) ensuring that both source and background data were available.
Lightcurves were background-subtracted using the instrumental
background, and deadtime corrected using {\sc hxfdtcor}. DYE$_{-}$ filtering is not carried out and so does not
affect the number of bursts seen; however, COR $>$ 6 is chosen, so the first burst seen in XIS is removed, but the
remaining bursts are seen.

%HXD data were collected during epoch 3 when High Voltage on PIN instrument
%was set as 400V/400V/500V/500V. XIS instrument was on-axis during the 
%observation. 

\section{Results}

\subsection{The burst rate}

The lightcurve of Fig. 1 shows 5 bursts, the last of which is double. We first obtain the mean time between 
bursts $\Delta t$ and thus the burst rate ($\Delta t^{-1}$). This was done by examining the XIS0 lightcurve with 
its full 2 s resolution, i.e. without 64 s binning. The times of the peaks of the bursts and separations between the 
bursts are listed in Table 1. The gap between the 1st and 2nd bursts is consistent with being
twice the approximate separation of bursts, and it is almost certain that one burst between these fell in a data gap 
and so we use half the gap between these bursts. We exclude the double burst from the analysis and using the gaps 
between the other bursts we find that $\Delta t$ = 20254 $\pm$ 106 s (with 1$\sigma$ errors), i.e.  
$\Delta t$ = 5.63 $\pm$0.03 hr.

Figure 2 (left panel) shows the evolution of burst recurrence time $\Delta t$ with time, i.e. MJD spanning 
the period 1985 to the {\it Suzaku} observation. Analysis producing $\Delta t$ is taken from  
Ba\l uci\'nska-Church et al. (2008) including values derived for the observations with {\it Exosat}, 
{\it ASCA}, {\it RXTE}, {\it XMM-Newton} and added to it is the point for the present {\it Suzaku} observation. 
It can be seen that the burst rate shows a systematic linear decrease from the {\it Exosat} observation to the 
{\it XMM} observation (2003, Jan 29) as shown by the best-fit solid line. The observations of {\it XTE} 2003 
(2003, Sep 25) and the present observation do not follow the trend as marked with the 
dotted line. It appears that a marked change took place within the source in 2003.

From spectral fitting below using the best fit to the broadband spectrum of the non-burst, non-dip data we 
obtain a 1 - 10 keV luminosity of $1.47\times 10^{36}$ erg s$^{-1}$, and we next compare the {\it Suzaku}
data with our previous work on XB\th 1323-619 in which we obtained the dependence of burst rate on source
luminosity from the {\it Exosat} observation of 1985 to the {\it RXTE} observation of 2003, a span of 18 years 
(Ba\l uci\'nska-Church et al. 2008). Burst rate as a function of luminosity is shown in \hbox{Fig. 2 (right panel)} 
including our analysis of the observations with {\it Exosat}, {\it ASCA}, {\it RXTE}, {\it XMM-Newton} and the present 
{\it Suzaku} observation. The solid line represents a linear fit to the data up to the 2003 {\it XTE} 
observations. Firstly, it can be seen that the 2003 {\it XTE} point follows the linear relation between the 
luminosity and the burst rate. Moreover, the point for {\it Suzaku} is close to the {\it Exosat} point, 
indicating not only that the source has jumped back to its luminosity of 1985, but also that the burst 
rate is consistent with that luminosity.

\subsection{The burst properties}

We determined the times of the burst peaks, the time between bursts and examined the decays of the
bursts and for this, the 64 s binning of Fig. 1 was not adequate and so we produced a lightcurve
from XIS0 with 2 s binning, this having more than sufficient count statistics. 
The decay profiles of the bursts were examined, and it was immediately obvious that each burst had a fast
decay followed by a much slower decay giving each burst a duration of more than 200 s. This behaviour was typical
of the source in previous observations (Ba\l uci\'nska-Church et al. 2008) and the slow decay indicates
nuclear burning by the the rp-process (Wallace \& Woosley 1981; Galloway et al. 2004) was operating.
We fitted each burst separately with a function 
$f(t) = A + N_1 \times exp[-(t-t_0)/\tau_1] + N_2 \times exp[-(t-t_0)/\tau_2$ in the vicinity of
each burst, i.e. extending 600 s beyond the burst peak. The function contains two time constants $\tau_1$ and $\tau_2$,
$t_0$ is the burst peak time, $A$ is the persistent emission count rate and  $\rm N_1$, $\rm N_2$ are normalizations.
The results are shown in Table 1, from which we find that the mean exponential decay times are $\tau_1$ = 9 $\pm$1.8 and
$\tau_2$ = 33$\pm$7 seconds. 
%The $\alpha$-parameter: the ratio of the non-burst fluence to the burst fluence was 
%also investigated. Earlier observations 
%having consistently a value of $\sim$ 45 indicating mixed H/He burning. However, in the {\it XMM}
%observation a jump to $\alpha$ $\sim$ 100 - 150 took place showing an increased contribution of He burni\ng.
%In the 2003 {\it RXTE} observation the changes continued such that it was not possible to detect a long
%tail in the burst decay which was characterized by a single time constant, suggesting less rp-burning.

\tabcolsep 1.2 mm
\begin{table}[!hb]
\begin{center}
\caption{Measured burst properties}
\begin{minipage}{80mm}
\begin{tabular}{lrrrr}
\hline \hline\\
Burst No. & Peak $\; \; \; \; \; \; \; $ & Separation & 
$\tau _1 $ & $\tau _2 $ \\
& Time (s) $^{(1)}$ & $\Delta$t (s) & (s) & (s)  \\
\noalign{\smallskip\hrule\smallskip}
\\
1 &  4148.0   & \dots       &$6.8 ^{+ 2.3}_{-1.5 }$  & $26.7^{+3.6}_{-2.4}$\\
\\
2 & 44358.0   & 40,210.0   &$10.4 \pm 2.8$          & $27.9^{+3.9}_{-8.9}$\\
\\
3 & 64696.0   & 20,338.0   &$6.7^{+2.9 }_{-2.0 }$   & $28.2^{+4.4}_{-2.7}$\\
\\
4 & 85016.0   & 20,320.0   &$11.1 \pm 1.2 $         & $46.1^{+28}_{-12.2}$\\
\\
5a & 107420.0 & 22,404.0   &$8.6 ^{+3.2}_{-2.7}$    & $35.5^{+4.6}_{-3.0}$\\
\\
5b & 108054.0 & 634.0      &$10.8 \pm 0.9$          &$^{(2)}$\\
\\
\noalign{\smallskip}\hline
\end{tabular}\\
$^{(1)}$  times relative to MJD 54109.48056.

$^{(2)}$ No long decay was present in burst 5b
\end{minipage}
\end{center}
\end{table}

The double burst is shown in the Table as bursts 5a and 5b; it is interesting that the second weaker
component 5b displayed no evidence for a slow decay.
 
\begin{figure}[!ht]                                                                 % Fig. 3
\begin{center}
\includegraphics[width=78mm,height=78mm,angle=270]{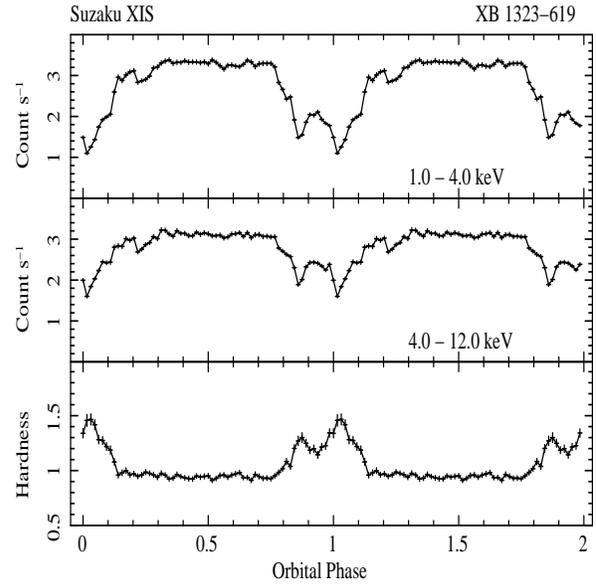}                        % was folded_new.qdp
\caption{Background subtracted XIS lightcurves (the sum of XIS0, XIS1 and XIS3)
with bursts removed folded on the derived orbital period of 2.928 hr using 64 phasebins per cycle:
upper panel: in the band
1.0 - 4.0 keV; middle panel: in the band 4.0 - 12.0 keV. The lower panel is the hardness ratio
formed by dividing the lightcurves in the higher and lower energy bands}
% lc_xi0_35m_ff_nobursts_folded_cts.qdp
\end{center}
\end{figure}

\subsection{X-ray dipping}

The X-ray dipping was examined by first creating a lightcurve from which all portions of all bursts were 
excluded to prevent this interfering with the determination of the best-fit orbital period using the dipping.
The XIS lightcurve 
which was a sum of background-subtracted lightcurves from XIS0, XIS1, XIS3 in the full band 0.2 - 12.0 keV 
was searched for periodicities using the {\sc period} program which gave a best fit period of 2.926 
$^{+0.38}_{-0.25}$ hr. To reduce the uncertainties, a second method was used in which the time of each dip 
minimum was obtained was plotted against cycle number and these data fitted. In the present observation,
the dips consisted of a first minimum followed by a second minimum within the envelope of the full dip, 
so the lowest point in the second part was used, there being 8 dips in which the time of this
was well-defined. This procedure produced a period of 2.928$\pm$0.002 hr. This is in good agreement with the 
previous determinations of 2.938$\pm$0.020 hr from {\it BeppoSAX} (Ba\l uci\'nska-Church et al. 1999) and
from {\it Exosat} of 2.932$\pm$0.005 hr (Parmar et al. 1989).
The data could then be folded on this period and in Fig. 3 we show 
the XIS lightcurves in two bands: 1.0 - 4.0 keV and 4.0 - 12.0 keV  and the hardness ratio formed by dividing 
these. The lightcurve was folded at the epoch of 54109.577080 MJD, which is the time of the first deep dip. 
Dipping is 55\% deep in the lower energy band and 36\% deep in the higher energy band in the folded light 
curve. However, folding introduces some smoothing and in the normal lightcurve (Fig. 1) dipping is of the order 
of 65\% deep.  The hardness ratio in Fig. 2 shows that lower energies are removed in dipping leading to the 
observed hardening of the spectrum, which is quite typical for the source. 

Similarly, PIN lightcurves extracted in various bands were examined and weak dipping with a depth of 
10\% $\pm$ 6 \% can be detected in a  folded lightcurve at energies up to 20 keV. At higher energies dipping 
was insignificant.

\subsection{The non-dip non-burst spectrum}

A non-dip, non-burst spectrum was accumulated from the whole observation by firstly noting the times of the  
start and end of bursts and removing the bursts by selecting on time. To remove 
dipping intervals and ensure that the 
shoulders of dipping were completely removed a selection on phase was made, phases -0.28 to +0.28 were removed 
where the zero of phase was defined as the deepest point of dipping.  A spectrum was selected from each of 
the three XIS detectors; the spectrum in each detector has an exposure time of about 34 ksec giving about 
84\th 000 counts in each detector (a count rate of 2.5 count s$^{-1}$). A background spectrum was accumulated 
with the same selections for each of the XIS instruments. A response file was generated using the 
{\sc ftool} {\sc xisrmfgen} and a corresponding auxiliary response 
file also generated for each  detector, using the latest calibration database of 18th April, 2008.
The {\it ASCA} ftool {\sc addascaspec} was then used to add the three spectra, the three corresponding 
background files, the three auxiliary response (arf) files and the three response matrix (rmf) files. 

The PIN spectrum was produced using the same GTI files as used to select the XIS data.
A burst-like event was noted in the background subtracted PIN lightcurve which occurred $\sim $900 s 
before burst no. 2 and lasted \hbox{$\sim $18 s.}  No increase of the count rate was 
detected in any of the three XIS instruments at this time. However, the time interval covering this 
event was removed from the selection of the PIN persistent emission spectrum. A background spectrum
was extracted from the non X-ray background events file provided for the observation, using the 
same GTI as above. The cosmic X-ray background was added as a term to the
spectral model,  with photon index  $\Gamma = 1.29$, a high energy cut-off at 40 keV  and
normalization of $\rm 8.8 \times 10^{-4}\; ph\; cm^{-2}\; s^{-1}\; keV^{-1}$ as required for
the observation with the source at the XIS aimpoint.

Deadtime correction of the source spectra was performed using {\sc hxddtcor}.
However, this ftool works correctly only good time intervals longer than 128 seconds.
Our PIN spectrum which was selected on orbital phase and for removal of bursts had a  tendency 
of having GTIs in quite short blocks of time, e.g. 16 s. This problems was solved using the standard
technique for correcting PIN lightcurves, i.e. by making a lightcurve with 128 s binning, selecting parts 
of this with GTIs required and deadtime corrected by extracting from the ``pseudo event file''
provided. Systematic errors of 3\% were added to all PIN spectra; no errors were added to XIS 
as is normal practice.

The XIS spectrum was fitted simultaneously with the PIN spectrum, parameter values being chained 
between the two. A multiplying factor was included in the model to allow for uncertainty in the relative 
instrument calibrations and was set to unity for the XIS and free for PIN as is usual practice.
Channels below 0.8 keV and above 12 keV in the XIS were ignored where the instrument response is not so 
well-determined; grouping to a minimum of 100 counts per channel was applied. In the PIN channels between 
12 and 70 keV were used, the instrument response at 10 - 12 keV being less certain and above 70 keV the
background was becoming large. 

A number of models were tried. Firstly, a cut-off power law  gave an acceptable fit with $\chi^2$/d.o.f. = 
1415/1633. However, since the 1990s it has generally been accepted that LMXB require a non-thermal 
and a thermal emission component as published evidence is very strong. White et al. (1988)
demonstrated the dominance of the Comptonized emission. Curvature in the continuum spectra at a few keV 
cannot be explained as a Comptonization high energy cut-off and requires the presence of a blackbody
(Church \& Balucinska-Church 2001), and as said in Sect. 1, the explanation of rapid variability in the dip sources
in dipping supports the presence of neutron star blackbody emission. Two-component models are clearly needed
to fit the most luminous LMXB (e.g. Schulz et al. 1989; Hasinger et al. 1990) and fitting two-component
models became general (e.g. Barret et al. 2000). Consequently, we next added a
simple blackbody to the spectral model although in faint sources like XB\th 1323-619 the thermal 
component is weak. This two-component continuum model consisting of an absorbed blackbody and cut-off power 
law ({\sc ag}$\ast${\sc (cut + bb)}) gave a significantly better fit for which an 
F-test showed that it was better with 99.9\% probability. Moreover, the normalization of the 
blackbody term was found to be significant at 2.6 $\sigma$,
%inconsistent with zero at 90\% confidence, 
so from this point we will 
assume the two-component model.
However, residuals were present indicating possible emission and several absorption features. 
Because of the decreased luminosity of the source compared with 
that at the time of the {\it XMM} observation, less line features were positively detected. There was some 
indication of a weak, broad emission line at $\sim$6.4 keV; however as discussed below, we regard detection
of this as only tentative.  The residuals also indicated an absorption feature at $\sim$6.7 keV and possible
absorption features at $\sim$6.6 and $\sim$6.9 keV as shown in Fig. 5. Finally, a trace of a possible absorption
at $\sim$9 keV can be seen in Fig. 5.
% Schulz cites 6.71 keV is He-like Fe XXV

Before adding lines to the spectral model we re-fitted the continuum model excluding the energy range
containing the lines 6.5 - 8.0 keV to prevent the lines slightly distorting the continuum fit. The continuum 
parameters were then frozen and lines added. The width of the emission line was fixed (at 0.02 keV)
to prevent the line absorbing neighbouring continuum.  Although the emission feature at $\sim$6.4 keV had
an equivalent width
EW of 17$\pm$7 eV, we do not include it in Table 2 showing the fit results as we regard the detection as tentative as the
appearance of the line in the residuals is not totally convincing. Initially a single absorption line was added 
with $\sigma$ fixed at 0.01 keV as all the absorption features appeared narrow. This fitting resulted
in a line at 
\begin{figure}[!ht]                                                    % Fig. 4
\begin{center}
\includegraphics[width=78mm,height=78mm,angle=270]{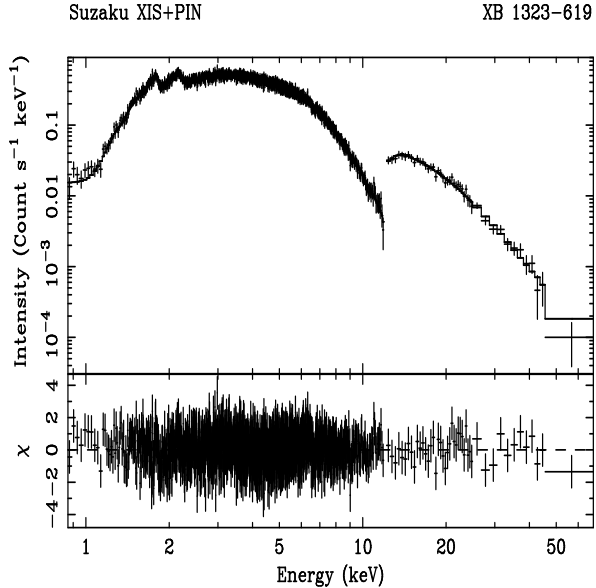}
                                                     % was fig4.qdp: model mjc55_noCXB.mod
\caption{Best fit to non-dip, non-burst spectrum of the three XIS detectors fitted simultaneously
with the PIN detector; for details of the line features detected see Fig. 5 and the text}
\end{center}
\end{figure}
6.6(7)$\pm$0.05 keV being detected with EW = 12$\pm$7 eV, and these results are shown in Table 2.
This line can be identified with Fe XXV and so represents absorption by highly ionized material
which could be located in the ADC. 

However, the residuals indicate there are probably at least three weak absorption lines and we attempted 
to fit the spectrum with three lines. In this case, it was not possible to determine the energies 
by free fitting so these were fixed at values obtained by inspection of the residuals, i.e. 6.59,
6.74 keV and 6.90 keV. 
It was not possible to get a sensible fit to the residual feature at 6.9 keV, i.e. the normalization
was zero. The other two lines at $\sim$6.59 and $\sim$6.74 keV were found to have EWs of 
-6.8$\pm$6.8 eV and -6.4$^{+6.4}_{-7.4}$ eV and correspond to marginal 1$\sigma$ detections, 
and for this reason we do not show them in Table 2. As said above, the energies are uncertain and 
we should be careful not to over-interpret them. 
%In fact, 6.59 keV if taken literally would correspond to a forbidden Fe XXV line. 
The feature at 6.59 keV may correspond to a less ionized absorber, 
that is, to an Fe K transition above about Fe XVII. As we see Fe XXV (Table 2) and also
absorption by neutral absorber in dipping, absorption by an intermediate ionization
state may be quite possible.
In Fig. 4 we show the folded data and residuals for the full band of the XIS and PIN
instruments, while in Fig. 5 (upper panel) we show the absorption features in the residuals
over the band 5 - 10 keV before lines were added to the spectral model. 

% NOTE: in the Table 
% continuum values change from previous version when we ignored 6.5 - 8.0 keV
% we show fits inclusing the emission line (even though this is not shown)
% it makes little difference to the absorption line fitting and non to the continuum which
% was obtained without any line.
% Fig. 4 left has residuals (top) without lines, bottom (with 1 emission line and 3 abs lines)
% Fig. 4 (right) has no emission line and abs lines make no differnce to figure.

In the best-fit solution, the best-fit power law photon index $\Gamma$ of  $1.67^{+0.10}_{-0.03}$ is very similar to 
that we obtained in the {\it XMM} observation of 1.68$\pm$0.08 (Church et al. 2005) and is physically reasonable; 
the cut-off energy of 85 keV is high as expected in a low luminosity source (a value of 44.1$^{+5.1}_{-4.4}$ 
was obtained in the broadband {\it SAX} observation (Ba\l uci\'nska-Church  et al. 1999). In the analysis of 
the {\it XMM} observation, absorption at 6.70$\pm$0.03 keV was seen and 6.98$\pm$0.04 keV, and broad emission 
at about 6.8 keV (Church et al. 2005).  Barnard et al. (2001) detected an emission line at 6.43$\pm$0.21 keV 
in a {\it RXTE} observation of XB\th 1323-619.

\subsection{The broadband deep dip spectrum}

It is, of course, of interest to investigate the spectrum of deep dipping, to 
use the broadband capacity of {\it Suzaku} to reveal the energy dependence of 
dipping, and test whether 
\begin{figure}[!ht]                                                    % Fig. 5
\begin{center}
\includegraphics[width=37mm,height=78mm,angle=270]{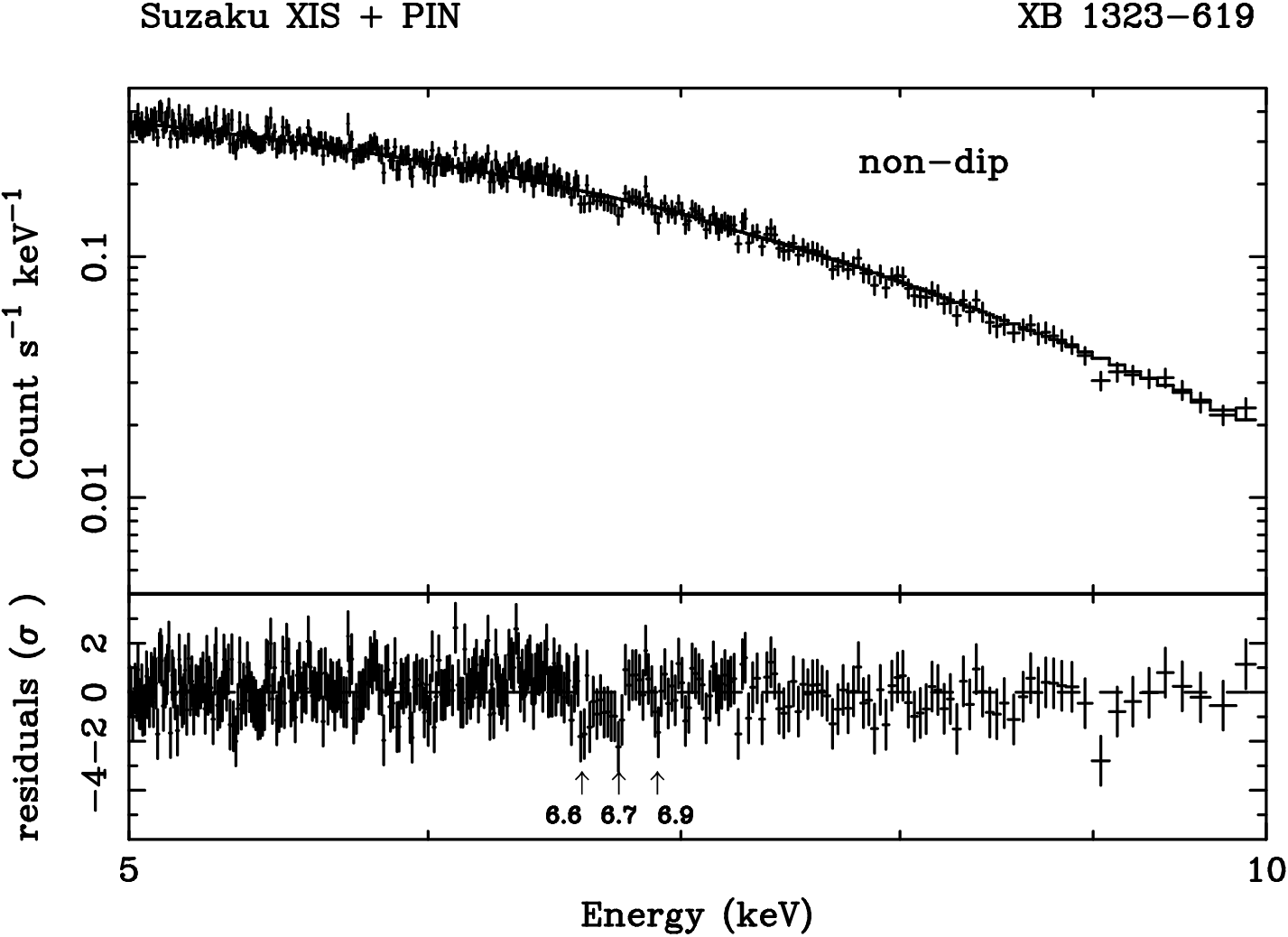}          % was fig5_pers_noCXB_res
\includegraphics[width=37mm,height=78mm,angle=270]{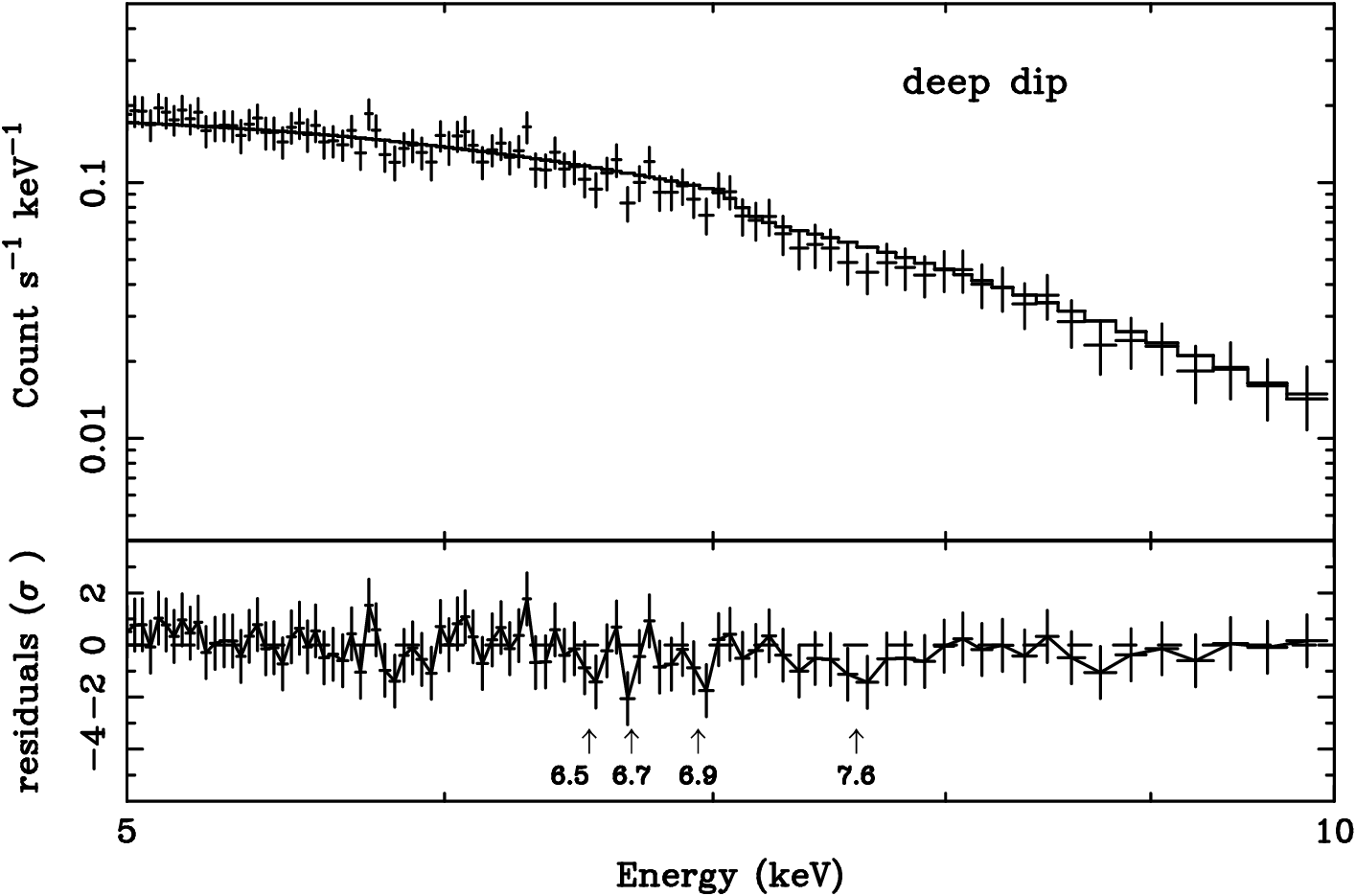}          % was fig5_dip_noCXB_res
\caption{Upper panel: persistent emission spectrum and model fit in the band 5 - 10 keV 
before addition of any line features showing several absorption features; lower panel: the 
corresponding residuals for the deep dip spectrum} 
\end{center}
\end{figure}
we can model the spectral evolution. In particular, 
we apply the ``progressive covering'' model for dipping (Church et al. 1997) in
 which the extended absorbing bulge on the outer accretion disk progressively 
overlaps the extended ADC source of Comptonized emission, this model having 
previously provided a very good description of spectral evolution in dipping in 
all dipping LMXB (Church et al. 1997).
We will also test the extent to which dipping can be described as photoelectric
absorption, and the degree to which electron scattering may take place, given
the suggestion by Boirin et al. (2005) based on analysis of the {\it XMM} 
observation of XB\th 1323-619 that spectral evolution in dipping in all the 
LMXB dipping sources might be explained by photoelectric absorption plus electron 
scattering  in a medium of varying ionization parameter.

Dip data were selected from the deepest part of dipping where dipping is $\sim$65\% deep, by 
selecting for intensities below 1.2 count s$^{-1}$ in XIS0 and applying the resulting GTI
\tabcolsep 3.5 mm
\begin{table*}
\begin{center}
\caption{Best simultaneous fit to the non-dip and dip XIS + PIN spectra.}
%\begin{minipage}{160mm}
\begin{tabular}{lrrr}
\hline \hline\\
Model Parameters & Non-dip & Dip\\
%model parameters & mjc51.mod & mjc50.mod & mjc53.mod \\
\noalign{\smallskip\hrule\smallskip}
$N_{\rm H}$ (10$^{22}$ atom cm$^{-2}$) & 3.2$\pm$0.1 &3.2$\pm$0.1\\
\\
{\sc cutoff power law} & & \\
$\Delta N_{\rm H}$ (10$^{22}$ atom cm$^{-2}$) &  0.0 & 22.2$\pm$0.8\\
Covering fraction $f$ & 0.0 & 0.63$\pm$0.02\\
photon index $\Gamma $ &$1.67^{+0.10}_{-0.03}$ \\
cut-off energy (keV) & $85^{+77}_{-35}$\\
normalization (10$^{-2}$ ph cm$^{-2}$ s$^{-1}$) & 2.4$^{+0.9}_{-0.3}$\\
\\
{\sc blackbody} & & \\
$\Delta N_{\rm H}$ (10$^{22}$ atom cm$^{-2}$) & 0.0 & $>$ 400\\ 
temperature $kT_{\rm BB}$ (keV) &$1.35\pm0.36$\\ 
normalization (10$^{-5}$ erg s$^{-1}$ at 10 kpc) & 8.3$\pm$5.1\\
\\
{\sc absorption line} & &  \\
energy $E$ (keV)&& 6.52$\pm$0.10 \\
width $\sigma $ (keV) && 0.01 (frozen) \\
normalization (10$^{-5}$ ph cm$^{-2}$ s$^{-1}$) &&-1.0$\pm$0.8 \\
equivalent width $EW$ (eV) && -18$\pm$15  \\
\\
{\sc absorption line} & &  \\
energy $E$ (keV)& 6.67$\pm$0.05 & 6.68$^{+0.18}_{-0.02}$\\
width $\sigma $ (keV) & 0.01 (frozen) & 0.01 (frozen)\\
normalization (10$^{-5}$ ph cm$^{-2}$ s$^{-1}$) &-1.24$\pm$0.71    & -0.9$^{+0.8}_{-0.9}$ \\
equivalent width $EW$ (eV) & -12$\pm$7  & -15$\pm$15\\
\\
{\sc absorption line} & &  \\
energy $E$ (keV)&& 6.94$\pm$0.14 \\
width $\sigma $ (keV) && 0.01 (frozen) \\
normalization (10$^{-5}$ ph cm$^{-2}$ s$^{-1}$)&&-1.5$\pm$1.4 \\
equivalent width $EW$ (eV) && -26$\pm$25 \\
\\
{\sc absorption line} & &  \\
energy $E$ (keV)&& 7.6$\pm$0.2\\
width $\sigma $ (keV) && 0.01 (frozen) \\
normalization (10$^{-5}$ ph cm$^{-2}$ s$^{-1}$) &&-2.8$\pm$2.1\\
equivalent width $EW$ (eV) && -68$\pm$52 \\
\\
dip/non-dip intensity shift && 0.79$\pm$0.04\\
\\
PIN normalization factor: & $1.12\pm 0.07$ & 1.12 (frozen)\\
\\
$\chi ^2 \; / d.o.f.$ & 1381/1634 & 140/263\\
$\chi ^2_{\rm r}$ & 0.85 & 0.53\\
\noalign{\smallskip}\hline
\end{tabular}\\
\vskip 1 mm
90\% confidence errors are given for all quantities except for the weaker 6.52\\ 
and 6.68 keV lines where 67\% confidence errors are shown for all parameters.
%\end{minipage}
\end{center}
\end{table*}
to all instruments. In addition, we removed bursts and selected for phases betweem -0.16 and +0.09.
Firstly, we fitted the XIS and PIN dip spectra with a model of the form 
{\sc ag}$\ast$({\sc pcf}$\ast${\sc cut} + {\sc ag}$_{1}$$\ast${\sc bb)} i.e. applying the
covering factor {\sc pcf} to the cut-off power law modelling the extended emission of the ADC, while the
point-like blackbody is subject to a column density {\sc ag}$_{1}$ in addition to the Galactic column density
{\sc ag}, frozen at the non-dip value. An assumption of this model is that 
emission parameters cannot vary in dipping, only absorption parameters and so all of the continuum emission 
parameters of the blackbody and cut-off power law were frozen at their non-dip values. 

Acceptable fits were obtained, with $\chi^2$/d.o.f. $\sim$ 1 but with clear residuals indicating the presence
of absorption features. The emission line seen at $\sim$6.4 keV in non-dip emission could not be seen in the 
residuals, and attempts to fit such a line resulted in a value of zero for the line normalization.
Consequently no emission line was included in dip fitting. Inspection of the residuals revealed absorption 
lines at $\sim$ 6.5, $\sim$ 6.7, $\sim$ 6.9 and $\sim$7.6 keV.
Corresponding lines were added to the spectral model. The line at 7.6 keV appeared broader
than the other lines in the residuals with a full width at zero height of $\sim$0.6 keV and so its width
$\sigma$ was initially set at a corresponding 0.15 keV in the fitting, while 
the widths $\sigma$ of the first three lines were frozen at a value
appropriate to narrow lines (0.01 keV).

A formally acceptable fit could be obtained in this way. However, a comparison of the non-dip and deep dip spectra
shown in Fig. 6 shows a vertically downwards shift in the dip spectrum, i.e. an energy-independent 
decrease of intensity.
%Including the lines, a good fit was obtained having $\chi^2$/d.o.f. of 164/263, a progressive covering fraction 
%$f$ = 0.70$\pm$0.01 and associated column density $N_{\rm H}$ = $33\pm 2\times 10^{22}$ atom cm$^{-2}$ 
%responsible for removing the cut-off power law, plus extra absorption of the blackbody $N_{\rm H}$ 
%$>$ $400\times 10^{22}$ atom cm$^{-2}$. Thus dipping is well-described by a large increase of column density 
%for the blackbody and a smaller $N_{\rm H}$ increase for the Comptonized emission as previously found 
%(e.g. Church et al. 1997) indicating that this emission is subject to absorption averaged across the bulge
%having density gradients. The covering fraction of 73\% shows that in deepest dipping, not all
%of the extended emission is overlapped by absorber, as shown by the depth of dipping in the lightcurve which
%does not exceed $\sim$65\%. It was possible to obtain the absorption line energies by free fitting
%and these were found to be 6.5$\pm$0.1, 6.7$\pm$0.1, 6.94$\pm$0.06 and 7.6$\pm$0.17 keV. However, the 
%XIS-PIN normalization factor was found to be 0.93$\pm$0.06 which is appreciably less than the non-dip 
%value of 1.12$\pm$0.07 and also less than values typically found, which indicates that the fit is not optimum.
%
%A comparison of the deep dip and non-dip spectra (Fig. 6) suggested a vertical downwards shift in the 
%deep dip spectrum, i.e. a possible energy-independent decrease of intensity. 
We investigated this, and
determined the level of any energy-independent change during dipping by adding a multiplying factor 
to the spectral model, this factor affecting both cut-off power law and blackbody terms. All emission parameters
were fixed at the non-dip values and the XIS-PIN normalization factor was fixed at the 
non-dip value of 1.12. There was a substantial improvement in the quality of fit with $\chi^2$/d.o.f. 
\begin{figure}
\begin{center}                                                                        % Fig. 6
\includegraphics[width=78mm,height=78mm,angle=270]{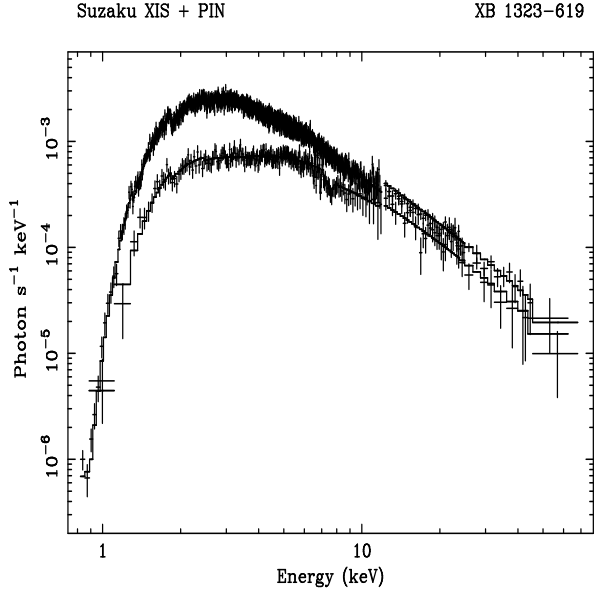}                          % was fig6_dnd_noCXB
\caption{Comparison of non-dip and deep dip spectra from best fits to XIS and PIN simultaneously
in each. For clarity individual model components are not shown to demonstrate the energy-dependent
nature of the dipping}
\end{center}
\end{figure}
becoming 140/263 compared with the value without the vertical shift factor (164/263).
Best-fit values for the final fit to the dip spectrum are shown in Table 2.
The dipping was well explained by an increase in column density ($\Delta$N$_{\rm H}$ 
of at least 400$\times 10^{22}$ atom cm$^{-2}$
and covering of 63\% of the Comptonized emission by absorber, with a smaller increase of $N_{\rm H}$ 
of $22\times 10^{22}$ atom cm$^{-2}$. This suggests that the blackbody is covered by central, denser regions
of the absorbing bulge whereas the Comptonized emission is overlapped by lower density regions on average.
Note that the best fit requires an unabsorbed fraction of 37\%, i.e. part of the spectrum not overlapped
is unabsorbed. In addition to the neutral absorption, the dip spectrum requires an energy-independent decrease 
by a factor of 0.79$\pm$0.04. 
It was found that the covering factor $f$ for the dip spectrum of 0.63$\pm$0.02 
agreed well with the depth of dipping in the lightcurve of $\sim$65\%.
Fig. 6 shows that below $\sim$20 keV dipping is strongly energy-dependent as modelled by neutral absorber in
our fitting. Above 20 keV the small, approximately energy-independent vertical shift can be seen.

%A test was made of including the emission line of the non-dip spectrum within the progressive covering factor
%of the dip model in order to determine whether the weakness, i.e. absence of this line, in dipping was consistent
%simply with its emission region being covered. For this test, the line parameters were fixed at the non-dip values.
%However, this fitting did not give the best-fit which was only achieved by freeing the line normalization which 
%then became zero. This gives an indication that the line does not originate in the full volume of the ADC
%for which the covering fraction determined applies, but in a smaller region and one which is more strongly
%absorbed in dipping.

Table 2 also shows the best-fit line energies, all obtained by free fitting.
The lines at 6.94 keV and 7.6 keV were detected at the 3-$\sigma$ level or higher, however, the other two
lines were weaker and were only detected at 1-$\sigma$ and so we show these lines in the Table with
1-$\sigma$ errors. Fig. 5 (lower panel) shows the absorption features using the best-fit continuum
model with the lines removed from the spectral model.
The line at $\sim$6.5 keV apparently corresponds to the weak feature seen in the non-dip spectrum
which we suggested was due to a less ionized Fe state. Possibly more than a single line is present,
i.e. a forest of lines from $\sim$ Fe XVII onwards. There appears no doubt that the line at
6.68 keV is Fe XXV as seen in non-dip. The equivalent widths of these lines are greater than
in non-dip; however, this may be due to the continuum decrease (EW = line flux/continuum intensity).
But because of the large uncertainties in EW values, we can only say that the increase in EW is 
consistent with the change in continuum. Indeed, it would not be expected that the bulge in the
outer disk would cause increased absorption in high ionization states.
The line at 6.94 keV, also seen in the observation with {\it XMM} (Church et al. 2005), can be
identified as Fe XXVI K$_{\alpha}$, and the line at 7.6 keV can be Ni XXVI.

%The additional absorption line at $7.6^{+1.4}_{-1.9}$ keV was best fitted by a broad Gaussian line 
%with $\sigma  1.2^{+2.9}_{-1.0}$ keV and EW of $0.4^{+1.3}_{-0.3}$ keV, which may suggest a blend of lines. 
%We also tried to fit the feature by an absorption edge but the fit did not allow this.

As a check, we also examined the energy-independent spectral change in the PIN spectrum alone, so as to be independent 
of XIS-PIN normalization using the energy band 20 - 70 keV where the effects of photoelectric absorption are 
small. We carried out spectral fitting in this band using a cut-off power law (as the blackbody flux is 
negligible) fitting the non-dip spectrum, and then fixing the power law index and cut-off energy in fitting the
dip spectrum. The normalization decreased by 24$\pm$3 \%, agreeing with the value of 21$\pm$4 \% obtained from 
simultaneous fitting of XIS and PIN; however this result may be biassed by points at 20 keV where absorption makes 
a small contribution and we conclude that our best value for the degree by which electron scattering reduces the
intensity is the factor shown in Table 3 from fitting both instruments simultaneously.

\section{Discussion}

Our spectral fitting results show that dipping consists of a large increase in photoelectric absorption
for the Comptonized emission of the ADC, the intrinsic column density increasing from zero to 
$22\times 10^{22}$ atom cm$^{-2}$ in deep dipping, this being consistent with covering of an extended ADC
source by an extended absorbing bulge in the outer disk as previously found for the dipping sources
in general, and for XB\th 1323-619 in the observations with {\it BeppoSAX} (Ba\l uci\'nska-Church et al. 1999),
with {\it RXTE} (Barnard et al. (2001) and {\it XMM} (Church et al. 2005). 
The neutron star blackbody is subject to a high column 
density which quickly removes this component in dipping as expected for a point-source and consistent with
the source being covered by the highest column density part of the bulge in the outer disk, rather than
an integration across the whole bulge as for the extended Comptonized emission.

An energy-independent shift can be seen at energies above 20 keV where photoelectric absorption is negligible 
indicating a degree of electron scattering, the shift being 21$\pm$4\%. A decrease of 10$\pm$10\% was found 
in our observation of the source with {\it BeppoSAX} in the band 20 - 50 keV (Ba\l uci\'nska-Church et al. 1999). 
We can show that the actual high energy shift is consistent with the degree of electron scattering expected 
for the electrons present in the bulge in the outer disk. At an energy of 5 keV, the photoelectric absorption 
cross section $\sigma_{\rm PE}$ is about 10 times the Thomson cross section $\sigma_{\rm T}$ and the column 
density for the Comptonized emission in dipping of $22\times 10^{22}$ atom cm$^{-2}$ reduces the intensity 
by a factor ${\rm exp}\,-N_{\rm H}\,\sigma_{\rm PE}$ $\sim$0.67 for the 63\% of the emission that is covered, 
so that the overall factor is 0.67$\times$0.63 + 0.37, i.e. a decrease of about 50\% at 5 keV close to the actual 
decrease as seen in Fig. 6. 

At 20 keV, $\sigma_{\rm PE}$ is 10 times less than $\sigma_{\rm T}$ and at higher energies becomes negligible. 
The electron density in the absorbing bulge in the outer disk will be between 1.0 and 1.2 times the ion
density depending on the exact ionization state, because of multiple electrons released in ionization
of the metals so that for an electron column density $N_{\rm e}$ of $\sim$22$\times 10^{22}$ atom cm$^{-2}$
electron scattering reduces the X-ray intensity of the covered part of the extended emission by a factor
0.86, so that the overall factor is 0.86$\times$0.63 + 0.37, i.e. 0.90, a 10\% reduction in intensity,
somewhat less than our measured decrease of 21\%, although the errors are substantial. Our spectral fitting was 
carried out using the cross sections for neutral absorber. We show below that the ionization parameter 
$\xi$ = $L/n\,r^2$, where $L$ is the luminosity, $n$ is the plasma density, and $r$ the radial position, can 
be no more than 50 in the outer disk because of the large distance from the neutron star. For $\xi$ smaller 
than $\sim$15, the degree of ionization in the bulge will be low, i.e. there will be a large neutral atom 
component. For $\xi$ between 15 - 50, only a low ionization state is achieved; for example, the oxygen ion 
may reach a state O\th II - O\th VI, but not the higher states O\th VIII or O\th IX (Kallman \& McCray 1982). 
The absorption cross sections are reduced from the neutral absorber values and in oxygen, $\sigma_{\rm PE}$ 
is reduced by up to a factor of $\sim$2 for O\th V / O\th VI (Verner et al. 1996). Thus spectral fitting 
would require a column density up to two times higher than for neutral absorber so that $N_{\rm e}$ would 
also be higher and the reduction in intensity by electron scattering would rise to 20\%. At higher energies, 
e.g. 20 - 70 keV, the Klein-Nishina formula shows that the mean Thomson cross section in this band decreases 
by 12\%, but this has little effect on the reduction in intensity so that, the expected decrease would again be 20\%.
%Klein-Nishina formula gives 5.865$\times 10^{-25}$ cm$^{-2}$, %(0.88 of 6.665)

Thus application of the progressive covering absorption model to the Extended ADC LMXB emission model
provides a very good description of the spectral evolution in dipping, including both the effects of
photoelectric absorption and electron scattering in the bulge in the outer disk. 
Assuming that the bulge has a radial extent of about 10\% of the disk radius
($\sim 3.5\times 10^{10}$ cm) and using the measured column density in dipping for the extended emitter 
(Table 3) we can estimate an average density in the bulge of $\sim 7\times 10^{13}$ cm$^{-3}$. From this
it follows that the ionization parameter $\xi$ cannot be larger than $\sim $50, so that the matter is 
in a relatively low ionization state ($\xi$ $\sim$ 1000 is needed for a high ionization state).
If our measured column densities are too low because we employ a neutral absorber model,
$\xi$ would be even smaller.
% but if N_H is x2 too low, n_e becomes 2x larger and xi even smaller.

Suggestions have been made by D\' iaz-Trigo et al. (2006) and Boirin et al. (2005) that dipping in all 
of the dipping LMXB, i.e. the large decreases in continuum flux, can be explained not in terms of absorption 
in the bulge in the outer disk but in an ionized absorber with high ionization state (log$\,\xi$ $>$ 3) suggesting that
the absorber is located much closer to the neutron star, although they do not specify the radial position 
of the absorber. 
%The only way it could be on the outer disk is if it is high above the orbital plane and the 
%density very low, but in this case, there would also be strong absorption in the bulge as normally accepted.
It is proposed from this fitting that a progressive covering description of dipping is not needed so that 
there is no need for an extended Comptonizing region. 
%{\bf However, in making these claims  
%work which presents strong evidence to the contrary, i.e. that the ADC is extended, is not cited 
%e.g. Church \& Balucinska-Church (2004)
%and Schulz et al. (2008a).  Moreover, very basic questions are not addressed: i.e. why does the
However, the questions of why the bulge in the outer disk does not cause strong absorption, and 
what azimuthal structure causes dipping in their model are not addressed.

Detailed recent work does reveal the presence of highly ionized line features in the dipping LMXB.
For example, Jimenez-Garate et al. (2003) investigated such features in XBT\th 0748-676
and concluded that dipping was due to neutral abosrber in the outer disk plus some contribution from ionized 
absorber also located in the outer disk but in layers above the disk. Parmar et al. (2002) report detection 
of broad emission and narrow absorption features in X\th 1624-490 using {\it XMM-Newton}. Lines present in 
dipping indicated that the additional absorbing material had lower ionization state than that producing the 
narrow absorption features seen in persistent emission.

However, the idea of a lower ionization state in dipping was taken by D\' iaz-Trigo et al. (2005) to mean
that a single region of high but changing ionization state could describe both non-dip and dip spectra,
without reference to the bulge in the outer disk. Spectral evolution in dipping in several sources was
modelled using a model {\sc abs}$\ast${\sc xabs}$\ast$({\sc bb+pl}) where {\sc abs} is neutral absorber and {\sc xabs} 
is ionized absorber. This model had free parameters $\xi$, ionized absorber column density 
$N_{\rm ion}$ and a line width $\sigma$.
The main parameter changes in dipping were a decrease of log$\,\xi$ from 3.5 typically to 3.0, an increase 
of $N_{\rm ion}$ by ten times, e.g. from $4\times 10^{22}$ to $40\times 10^{22}$ atom cm$^{-2}$ and a small increase 
in the column density of neutral absorber, e.g. typically by up to a factor of two from $0.3\times 10^{22}$ 
to $0.7\times 10^{22}$ atom cm$^{-2}$. In essence, this fitting still explains dipping as photoelectric absorption
but in a region where the net effective cross section of the mixture of elements is reduced as the
metals are in higher ionization states. The continuum change in dipping can be fitted, but the high $\xi$ has the 
consequence that the absorber has to be located much closer to the neutron star than the bulge.
% where it is
%difficult to understand what azimuthal structure could exist, unlike absorption in the bulge.

Thus D\' iaz-Trigo et al. (2006) based on this spectral fitting, and Boirin et al. (2005), conclude that
dipping can be explained without progressive covering of an extended emission region as applied
extensively to the dipping sources, e.g. by Church et al. (1997). They further suggest that an extended 
ADC is not required in their fitting, which is the basis of the Extended ADC emission model for LMXB in 
general (Church \& Ba\l uci\'nska-Church 2004) for which substantial evidence exists. 
Firstly, fitting non-dip and dip spectra in any dipping LMXB shows that the 
majority of the X-ray flux, i.e. the non-thermal emission, is removed gradually and systematically
proving that the emitter is extended (e.g. Church et al. 1997). The size of the Comptonization emission site
has been measured by dip ingress timing for all of the dipping LMXB (Church \& Ba\l uci\'nska-Church
2004) showing that this emitter is indeed very large, having a radial extent typically 50\th 000 km
or 5 - 50\% of the accretion disk radius. Strong independent evidence for an extended hot ADC is given by
Schulz et al. (2009) who detect a variety of broad emission lines of highly ionized species including
Ne\th X L$_{\alpha}$, Mg\th XII L$_{\alpha}$, Si\th XIV L$_{\alpha}$, S\th XVI L$_{\alpha}$,
 Fe\th XXIV, Fe\th XXV and Fe\th XXVI in {\it Chandra}
grating spectra of Cygnus\th X-2. The emissivities indicate log$\,\xi$ $>$ 3, and the radial
positions of the emitters were found to be between 20\th000 and 110\th 000 km from the  neutron star.
It was thus concluded that these lines originate in a hot, extended ADC and the radial distances
are in excellent agreement with the results of dip ingress timing. 
Schulz et al. (2008) in a study of Cir\th X-1 similarly found a large number of emission line features
rich in H-like and He-like features of high Z-elements such as Si, S, Ar and Ca. The spectra could be modelled
as emission from a hot ($T$ $\sim$ 10$^7$ K) photoionized plasma with $\xi$ = 1000, located at a radial
distance $r$ $\sim$ 10\th 000 km from the neutron star, i.e. a hot extended accretion disk corona.

Given the now apparent overwhelming evidence for the existence of extended ADCs, it now seems
necessary to include such an extended emitter in any model fitting (i.e. by a progressive covering term), 
which is quite in contrast to simply using an ionized absorber as done, for example, in Diaz-Trigo et al. (2006).
In addition, there are a number of elements lacking in the model which appear difficult to
explain in terms of the model, as follows, which are not a problem for the standard explanation of dipping.
It is not specified at what radial position the ionized absorber exists, nor is it suggested what causes the
azimuthal structure required in the absorber to cause dipping in a particular range of orbital
phases. It is 
%while the ionized absorber causing the narrow absorption lines that are independent
%of orbital phase (in non-dip emission) must have cylindrical symmetry, and it is 
difficult to see why there should be any azimuthal structure in an inner region where $\xi$ is high, especially
since such structure would have to be similar to the known structure in the outer disk, i.e.
the bulge. Finally the model does not explain why the bulge should have no absorbing effect
contrary to accepted ideas. Contrasted with this should be the existing
explanations of dipping as progressive covering of the neutron star and an extended ADC by
the extended absorbing bulge, which is physically reasonable, with substantial evidence for it.

From the continuum changes, we turn to the line features detected in the {\it Suzaku} observation
(Table 2). A possible emission line consistent with 6.4 keV was seen in the {\it Suzaku} non-dip 
spectrum which was absent in the dip spectra. A similar feature had been seen in the 1997 {\it RXTE} 
spectrum at 6.43$\pm$0.21 keV (Barnard et al. 2001) although the line was not found in the 2003 {\it XMM} 
observation when the source was much brighter. Absorption lines were also seen: in the non-dip spectrum
the only line we can be confident of was the 6.67 keV Fe XXV line. In dipping, four absorption lines 
were seen as discussed previously. The general presence of these medium to high ionization features
are consistent with formation in a hot ADC and this hot extended ADC is, of course, a major element
of the ``Extended ADC'' continuum emission model (Church \& Ba\l uci\'nska-Church 2004). Other authors,
in particular Schulz et al. (2008, 2009) have presented strong evidence that similar lines 
originate in an extended, hot ADC.

In the {\it Suzaku} observation of XB\th 1323-619 the luminosity was at a low level in contrast with
the highest luminosity exhibited by the source in the 2003 {\it XMM} observation, after which the 
luminosity but reduced by about a factor of two in the {\it RXTE} observation of 2003, and fell further
by almost another factor of two in the present observation. The X-ray burst rate has also decreased
substantially to about the value it had in the {\it Exosat} observation, as shown in Fig. 2.
The present burst rate supports the idea that the burst rate follows a definite burst rate-luminosity
relation which has not changed over 20 years.
We found in the present observation that the typical long
burst decays having two time constants indicative of burning by the rp-process. Such behaviour
was found previously in the source prior to the {\it XMM} observation. However, in the {\it XMM}
observation it wasd found that the long tails were not seen in bursts indicating less rp-burning
(Ba\l uci\'nska-Church et al. 2008), 
a possible reason being come complete burning of H to He between bursts at the higher luminosity
and mass accretion rate. 
%In {\it XMM} an increase of the $\alpha$ parameter was seen consistent
%with expectations since shorter bursts have smaller burst fluences. 
The return of the burst
properties of the source to those determined for the {\it Exosat}, {\it RXTE} and {\it BeppoSAX}
observations prior to the {\it XMM} observation support the idea that the burst behaviour is
determined primarily by the mass accretion rate.

The evidence is that the luminosity measured in the {\it Exosat} and {\it Suzaku} observations
is a lower limit for the source. Between 1985 and 2003, the luminosity increased by a factor 
of about 4, implying that the mass accretion rate increased by that factor. We cannot say whether 
this behaviour is periodic or not, however, the observed gradual increase over an 18 year
period followed by a much faster decrease in luminosity over 4 years does not suggest a periodic
change, and further observations will reveal whether the luminosity will increase as it did
from 1985 onwards. If the source were periodic, it could possibly be a triple system, in which
case the third body would need to have a period of the order of 30 days.

%There are sometimes long periodicities in LMXB such as a period of 176 days in
%4U\th 1820-303 (Priehorsky \& Terrell 1984). Zdziarski et al. (2007) proposed that this could be explained
%as 4U\th 1820-303 being a triple source; in this case they derive a period of the third body of
%a fraction of a day.
%
%In the case of XB\th 1323-619 we do not have evidence for periodicity, moreover the longterm
%variation in luminosity of the sources consists of a rather gradual growth, with increasing
%rate of increase up to 2003, i.e. over 18 years followed by a much faster decrease back to the 
%initial luminosity in 4 years. If we were to assume a triple system, the period of the third 
%body of 45 days assuming a modulation period of 40 years, i.e. twice the observed half-cycle.
%However, the very non-sinusoidal variation argues against this explanation. It is possible 
%that a completely different explanation must be found.

\thanks{We are grateful to the referee for his very helpful comments.
This work was supported in part by the Polish Ministry of Higher 
Education and Science grants KBN-1528/P03/2003/25 and 3946/B/H03/2008/34,
and by PPARC grant PP/C501884/1. We thank P. J. Humphrey for providing 
additional analysis tools.

\end{document}